\documentclass[prl,aps,twocolumn, showpacs]{revtex4}
\usepackage[dvips]{graphicx}
\usepackage{dcolumn}
\usepackage[english, russian]{babel} %- для рус-англ. переноса
\usepackage{amsmath, latexsym,  amssymb, array, graphics,
amsfonts, amsthm, amsmath,  amsfonts, array, bm, eucal}
\newcommand{\const}{\mathop{\rm const\, }}

{

\begin{document}
\newcommand{\mc}[1]{\mathcal{#1}}
\newcommand{\E}{\mc{E}}
\topmargin=-15mm
%   \Large

\title {\bf
Smoluсhowski problem for degenerate Bose gases}
%cond-mat/0601600

\author{\bf Anatoly V. Latyshev and Alexander A. Yushkanov}

\affiliation  {Department of Mathematical Analysis and
Department of Theoretical Physics, Moscow State Regional
University,  105005, Moscow, Radio st., 10--A}

\begin{abstract}
We construct a kinetic equation simulating the behavior of
degenerate quantum Bose gases with the collision rate
proportional to the molecule velocity. We obtain an analytic
solution of the half--space boundary--value Smoluchowski problem
of the temperature jump at the interface between the degenerate
Bose gas and the condensed phase.

\end{abstract}
%permeability - проницаемость
\date{\today}
\maketitle

\section{1. Introduction}

In recent years, the behavior of quantum gases has increased
interest. In particular, this is related to the development of
experimental procedure for producing and studying quantum gases
at extremely low temperatures  \cite{Pi}.
In the majority of papers, bulk properties of quantum gases have
been studied \cite{Spon} and \cite{Pang}.
At the same time, it is obvious that it is
important to take boundary effects on the properties of such
systems into account. In particular, such a phenomenon as the
temperature jump at the interface between a gas and a condensed
(in particular, solid) body in the presence of a heat flux
normal to the surface is important. Such a temperature jump is
frequently called the Kapitsa temperature jump \cite{Enz}.
We also mention a paper where the problem of boundary conditions
for the motion of superliquids was considered \cite{Pomeau}.

The problem of a temperature jump in a quantum Fermi gas was studied in
\cite{Lat1}, where an analytic solution for an arbitrary degree of gas
degeneracy was obtained. A similar problem for a Bose gas was
considered in \cite{Lat2}, where the gas was assumed to be
nondegenerate, i.e., it was assumed that there was no Bose --- Einstein
condensate.

The present paper is devoted to solving the problem of a
temperature jump in a degenerate Bose gas analytically. The
presence of a Bose --- Einstein condensate \cite{Lan}
leads to a considerable
modification of  both the problem statement and its solution
method. In this case, a kinetic equation with a model collision
integral is used to describe kinetic processes. We assume that
the boundary conditions at the surface are purely diffusive.

\section{2. Kinetic equation}

To describe the gas behavior, we use a kinetic equation with a
model collision integral analogous to that used to describe a
classical gas. In this case, we take into account the quantum
character of the Bose gas and the presence of the Bose --- Einstein
condensate.

For a rarefied Bose gas, the evolution of the molecule
distribution function $f$ can be described by the kinetic equation
$$
\dfrac{\partial f}{\partial t} +
\dfrac{\partial \E}{\partial {\bf p}}
\nabla f=I[f],
$$
where $\E$ is the kinetic energy of molecules, $\bf p$ is the molecule
momentum, and $I[f]$ is the collision integral.

In the case of the kinetic description of a degenerate Bose gas,
we must take into account that the properties of the Bose --- Einstein
condensate can change as functions of the space and time coordinates.
In other words, we must consider a two--liquid model
(more precisely, a two--fluid model, because we consider a gas
rather than a liquid). We let  $\rho_c=\rho_c({\bf r},t)$
and ${\bf u}={\bf u}({\bf r},t)$ denote
the respective density and velocity
of the Bose condensate --- Einstein.
We then have the expressions \cite{H}
$$
{\bf j}=\rho_c {\bf u},\qquad {\bf Q}=\dfrac{\rho_c u^2}{2}{\bf u},
$$
$$
\Pi_{ik}=\rho_c u_iu_k.
$$
for the densities ${\bf j}$ and ${\bf Q}$ of the respective mass and energy
fluxes and for the momentum flux tensor $\Pi_{ik}$
of the Bose condensate --- Einstein
(under the assumption that the chemical potential is zero).

The conservation laws for the number of particles, energy,
and momentum require that the relations
$$
\dfrac{\partial \rho_c}{\partial t}+\nabla {\bf j}=
-\int I[f]d\Omega_B,
$$
$$
\dfrac{\partial E_c}{\partial t}+\nabla {\bf Q}=
-\int \dfrac{p^2}{2m}I[f]d\Omega_B,
$$
$$
\dfrac{\partial (\rho_c{\bf u})}{\partial t}+\nabla {\Pi}=
-\int {\bf p}I[f]d\Omega_B
$$
be satisfied, where $s$ is the molecule spin,
$$
d\Omega_B=
\dfrac{(2s+1)d^3p}{(2\pi \hbar)^3}, \qquad
E_c=\dfrac{\rho_c u^2}{2},
$$
where $\hbar$ is the Planck constant.

In what follows, we are interested in the case of stationary
motion with small velocities (compared with the thermal velocities).
We note that for the Bose condensate, the quantities ${\bf Q}$  and
$\Pi_{ik}$, are
depend nonlinearly on the velocity (they are proportional to the
respective third and second powers of the velocity). Therefore,
in the approximation linear in the velocity ${\bf u}$,
the energy and momentum conservation laws can be written as
$$
\int \dfrac{p^2}{2m}I[f]d\Omega_B
$$
and
$$
\int {\bf p}I[f]d\Omega_B=0.
$$

According to the Bogolyubov theory,
the relation for the excitation energy $\varepsilon (p)$ \cite{Lan}
$$
\varepsilon (p)=\left[u^2p^2+
\left(\dfrac{p^2}{2m}\right)^2\right]^{1/2},
\eqno{(1)}
$$
where
$$
u=\left(\dfrac{4\pi  \hbar^2an}{m^2}\right)^{1/2}.
$$

Here $a$ is the scattering length for gas molecules, $n$ is the
concentration, $m$ is the mass, and $\bf p$ is the momentum of the gas
molecule, holds for a weakly interacting Bose gas. The parameter
$a$ characterizes the interaction force of gas
molecules and can be assumed to be small for a weakly interacting gas.

The relation $u^2\ll kT/m$, where $k$ is the Boltzmann constant
and $T$ is the gas temperature, holds for sufficiently small $a$.
In this case, we can neglect the first term in the brackets in (1).
The expression for the energy $\E(p)$ takes the
same form as in the case of noninteracting molecules:
$$
\E(p)=\frac{p^2}{2m}.
$$

We now consider the widely used kinetic equation in the
Boltzmann --- Krook --- Welander form with the molecule
collision rate proportional to the molecule velocity \cite{Lat1, Lat2, Cer}
$$
\dfrac{\partial f}{\partial t}+ ( {\bf v}\nabla )f=
\nu_0 w(f^*_M-f).
\eqno{(2)}
$$

Here, $f$ is the distribution function, $\bf v$ is the molecule velocity,
$w=|{\bf v}- {\bf v}_0|$, ${\bf v}_0$ is the mean--mass
gas velocity, $f^*_M$ is the Maxwell distribution function,
$$
f^*_M=n_*\left( \dfrac{m}{2 \pi kT_*}\right)^{3/2}
\exp \left[- \dfrac{m}{2kT_*}({\bf v}- {\bf u}_*)^2\right],
$$
and $\nu_0$  is a model parameter
corresponding to the inverse mean free path $l$ of a molecule,
$\nu_0 \sim 1/l$.

The parameters in the formula for $f_M^*$, namely,
$n_*$, $T_*$ and ${\bf u}_*$, can be determined from the
conservation laws for the number of molecules, momentum, and energy
$$
\int wf\,d\Omega_M =\int w f_M^* \,d\Omega_M ,
\eqno{(3a)}
$$
$$
\int w {\bf v}f\,d\Omega_M =\int w {\bf v}f_M^* \,d\Omega_M,
\eqno{(3b)}
$$
$$
\int w\dfrac{m}{2}({\bf v}-{\bf u})^2f\,d\Omega_M =
\int w \dfrac{m}{2}({\bf v}-{\bf u})^2f_M^* \,d\Omega_M,
\eqno{(3c)}
$$
where
$$d\Omega_M=d^3v.$$

The conservation law for the number of particles in the normal
state is inapplicable because the transition of particles to
the Bose --- Einstein condensate can occur. As mentioned above, the effect
of the condensate on the energy and momentum
conservation laws can be neglected in the approximation linear in
$\mathbf{u}$.

We note that Eq. (2) corresponds to the assumption that
the mean free path of molecules is constant
(it is independent of molecule velocities).
It hence follows that Eq. (2) to a greater extent corresponds to the
model in which the molecules are regarded as solid spheres.

We consider a generalization of Eq. (2) to the case of a
degenerate Bose gas. We assume that the general structure
of Eq. (2) is preserved, but by a function $f_M^*$, we must mean the
Bose  --- Einstein distribution (Bosean) with a zero chemical potential
\cite{Lan}
$$
f_B^*= \left[ \exp\left(\dfrac{m}{2kT_*}
({\bf v}-{\bf u}_*)^2\right)-1\right]^{-1}.
$$

Here, the parameters $T_*$ and ${\bf u}_*$ , are determined by the
second and third conditions in (3). In this case, we have
$$
d\Omega_B= \dfrac{2s+1}{(2 \pi \hbar)^{3}}d^3{p}
$$
instead $d\Omega_M$. %$\hbar$ is the Plank constant.

We assume that the mass velocity of the gas is much less
than the mean thermal velocity of the molecules and the
typical temperature variations along the mean free path
of molecules are small compared with the gas temperature.
The problem can be linearized under these assumptions.

We seek the distribution function in the form
$$
f=f^{s}_B(v)+ \varphi(t,{\bf r},{\bf v})g(v),
$$
where
$$
f^{s}_B(v)=\dfrac{1}{\exp(\beta_s v^2)-1},\qquad
\beta_s=\dfrac{m}{2kT_s},
$$
$\varphi$ is a new unknown function, $T_s$ is the surface temperature,
and
$$
g(v)=- \dfrac{\partial }{ \partial \varepsilon_s}f^{s}_B,\qquad
\varepsilon_s= \beta_s v^2.
$$

We introduce the notation
$$
{\bf C}=\sqrt{\beta_s}{\bf v},\qquad\;
\varepsilon_*=\dfrac{m}{2kT_*}({\bf v}-
{\bf u}_*)^2.
$$

Taking this notation into account, we have
$$
f^*_B(\varepsilon_*)= \dfrac{1}{\exp(\varepsilon_*)-1},
$$
$$
f^{s}_B(C)= \dfrac{1}{\exp(C^2)-1},
$$
$$
g(C)=\dfrac{\exp(C^2)}{(\exp(C^2)-1)^2}.
$$

We linearize the local Bose --- Einstein distribution  $f_B^*$,
passing to dimensionless quantities. We note that
$$
\varepsilon_*= \dfrac{T_s}{T_*}\left[\beta_s({\bf v}-{\bf
u}_*)^2\right]= \dfrac{T_s}{T_*}\left[({\bf C}- {\bf
W})^2\right],
$$
where
$$
{\bf W}=\sqrt{\beta_s}{\bf u}_*.
$$

Taking $T_*=T_s+ \delta T_s$ into account, we obtain
$$
\varepsilon_*=
C^2-\dfrac{ \delta T_*}{T_s}C^2-2 {\bf C}{\bf W},
$$
whence we find
$$
\delta
\varepsilon_*=-2 {\bf C}{\bf W}- C^2\dfrac{
\delta T_*}{T_s},
$$
where
$$
\delta \varepsilon_*= \varepsilon_*-
\varepsilon_s,\;\qquad \varepsilon_s=C^2.
$$

Consequently,
$$
f^*_B=f^{s}_B+ \left(\dfrac{ \partial f_B^*}{\partial
\varepsilon_*}\right)_{\varepsilon_*=\varepsilon_s}
\delta \varepsilon_*,
$$
or
$$
f^*_B=f^{s}_B+ g(C)\left[2 {\bf C} {\bf W}
+C^2\dfrac{\delta T_*}{T_s}\right].
$$

We note that the quantity $w$ in Eq. (2)
can be replaced with $v$ in the approximation
under consideration. We introduce the dimensionless
quantities  $t^*=t\nu_0/\beta_s$ and ${\bf r}^*={\bf r}\nu_0$
and omit the
asterisks on these quantities below. It is now clear that
Eq. (2) (in the dimensionless variables) becomes
$$
\dfrac{\partial \varphi}{\partial t}+({\bf C}\nabla)\varphi=
2C({\bf C}{\bf W})+C^3\dfrac{ \delta T_*}{T_s}- C \varphi.
\eqno{(4)}
$$

The parameters of this equation can be found from the momentum
and energy conservation laws (relations (3)), which now become
$$
\int (L \varphi)C^2 g(C)\,d^3C=0,
$$
$$
\int (L \varphi) {\bf C}g(C)\,d^3C=0,
$$
where
$$
L\varphi=2 C{\bf C}{\bf W}+C^3
\dfrac{\delta T_*}{T_s} -C\varphi.
$$

From this system, we find
$$
{\bf W}=\dfrac{3}{8\pi g_0}
\int \varphi \,{\bf C}\,g(C)\,C\,d^3C,
$$
$$
\dfrac{\delta T_*}{T_s}=
\dfrac{1}{4\pi g_2}\int \varphi g(C)C^3\,d^3C,
$$
where
$$
g_n=\int\limits_{0}^{\infty}C^{5+n}g(C)\,dC,\qquad n=0,1,2,
$$

In this case, we have
$$
g_0=\int\limits_{0}^{\infty}g(C)C^5\,dC=
\int\limits_{0}^{\infty}\dfrac{\exp(C^2)C^5\,dC}
{\Big(\exp(C^2)-1\Big)^2}=$$$$=-2\int\limits_{0}^{\infty}
C\ln(1-\exp(-C^2))\,dC=\dfrac{\pi^2}{6}=1.64493,
$$
$$
g_1=\int\limits_{0}^{\infty}g(C)C^6\,dC=2.22912,
$$
$$
g_2=\int\limits_{0}^{\infty}g(C)C^7\,dC=3.60617.
$$

We represent Eq. (4) in the standard form
$$
\dfrac{\partial \varphi}{\partial t}+({\bf C}\nabla)\varphi+
C\varphi(t,{\bf r},{\bf C})=
$$
$$=
\dfrac{C}{4\pi}\int k({\bf C},{\bf C'})\varphi(t, {\bf r},{\bf
C'})\,C'g(C')d^3C',%\Omega(\alpha),
\eqno{(5)}
$$
где
$$
k({\bf C},{\bf C'})=\dfrac{3}{g_0}{\bf C}{\bf C'}+
\dfrac{1}{g_2}C^2{C'}^2.
$$

\section{3. Problem statement}

A degenerate Bose gas occupies the half-space $x > 0$
above the plane surface in the problem under consideration.
A heat flux $Q$ normal to the surface is maintained in the gas.
We let $T_0$ denote the gas temperature far from the surface.
The quantity $T_0$ differs from the surface temperature
$T_s$ if
there is a heat flux. We let $\Delta Т = T_0 - T_s$ be the difference
between these temperatures. The quantity $\Delta Т$ is called the
temperature jump (the Kapitsa temperature jump in the case
of low temperatures).

The problem is to find $\Delta T$ as a function of the heat flux $Q$.
Taking into account that the problem is linear, we can write
$$
\Delta T=T_Q Q.
$$

The quantity $T_Q$ is called the coefficient of the
temperature jump. Another notation can be used:
$$
Q=R \Delta T,
$$
where $R$ is called the Kapitsa resistance.

Taking into account that the problem is stationary and that
the function $\varphi$ is
independent of the coordinates $у$ and $z$, we simplify Eq. (5):
$$
\mu \dfrac{\partial \varphi}{\partial x}+ \varphi(x,\mu,C)=$$$$=
\dfrac{1}{2}\int\limits_{-1}^{1} \int\limits_{0}^{\infty}k(\mu,C;\mu',C')
\varphi(x,\mu',C')\,d\Omega(C'),
\eqno{(6)}
$$
where
$$ \begin{array} {l}
\mu=\dfrac{C_x}{C},\qquad
d\Omega(C')= g(C') {C'}^3\,d\mu'dc', \\ \\
k(C,\mu; C',\mu')=\dfrac{3}{g_0}C\mu C'\mu'+\dfrac{1}{g_2}C^2{C'}^2.
\end{array}
$$

It is easy to verify that Eq. (6) has the particular solutions
$$
\varphi_1=\mu C,\qquad \varphi_2=C^2.
$$

Consequently, the function
$$
\varphi_{as}(x,\mu,C)=B\mu C+\varepsilon_T C^2,
\eqno{(7)}
$$
where $В$ is proportional to the heat flux,
is an asymptotic distribution function (as $x\to +\infty$).

Assuming that the reflection of the molecules from
the wall is purely diffusive,
we now formulate the boundary conditions
$$
\varphi(0,\mu,C)=0,\quad 0< \mu<1,
\eqno{(8)}
$$
$$
\varphi(x,\mu,C)=
$$
$$
=\varphi_{as}(x,\mu,C)+o(1),\;
x\to +\infty,\;-1< \mu<0.
\eqno{(9)}
$$

\section{4.  Reduction to the one--velocity problem and
the separation of variables}

We seek the solution of problem (6)--(9) in the form
$$
\varphi(x,\mu,C)=Ch_1(x,\mu)+C^2h_2(x,\mu).
\eqno{(10)}
$$

We obtain the system of equations
$$
\mu\dfrac{\partial h_1}{\partial x}+h_1(x,\mu)=
$$
$$
=\dfrac{3}{2}\mu
\int\limits_{-1}^{1}\mu' h_1(x,\mu')d\mu'+\dfrac{3g_1}{2g_0}\mu
\int\limits_{-1}^{1}\mu' h_2(x,\mu')\,d\mu',
$$
$$
\mu\dfrac{\partial h_2}{\partial x}+h_2(x,\mu)=
$$
$$
=\dfrac{g_1}{2g_2}\int\limits_{-1}^{1}h_1(x,\mu')d\mu'+
\int\limits_{-1}^{1}h_2(x,\mu')d\mu'.
$$

We represent this system of equations with respect to the column vector
$$
h(x,\mu)=\left[
  \begin{array}{c}
    h_1(x,\mu) \\h_2(x,\mu)  \end{array}\right]
$$
in the vector form
$$
\mu \dfrac{\partial h}{\partial x}+h(x,\mu)=\dfrac{1}{2}
\int\limits_{-1}^{1}
K(\mu,\mu')h(x,\mu')\,d\mu',
\eqno{(11)}
$$
where $K(\mu,\mu')$ is the kernel of Eq. (11),
$$
K(\mu,\mu')=
\left[\begin{array}{cc}3\mu\mu' & 3\dfrac{g_1}{g_0}\mu\mu'\\
\dfrac{g_1}{g_2}& 1
 \end{array}\right].
$$

Using (10), we transform boundary conditions (8) and (9) into
$$
h(0,\mu)={\bf 0}, \quad 0<\mu<1, \qquad {\bf 0}=\left[
  \begin{array}{c}
    0\\0  \end{array}\right],
\eqno{(12)}
$$
$$
h(x,\mu)=
$$
$$
=h_{as}(x,\mu)+o(1),\quad\;x\to +\infty,\quad\;-1<\mu<0,
\eqno{(13)}
$$
where $h_{as}(x,\mu)$ is the asymptotic distribution function
$$
h_{as}(x,\mu)=\left[\begin{array}{c}
B\mu\\
\varepsilon_T
\end{array}\right].
$$

We note that the equation kernel can be represented in the form
$$
K(\mu,\mu')=K_0+3\mu\mu'K_1,
$$
$$
K_0=\left[
  \begin{array}{cc}
    0 & 0 \\
    \dfrac{g_1}{g_2} & 1
  \end{array}\right], \quad
K_1=\left[
  \begin{array}{cc}
    1 & \dfrac{g_1}{g_0} \\
    0 & 0
  \end{array}\right].
$$

The general method for the separation of variables yields the expression
$$
h_\eta(x,\mu)=\exp(-\dfrac{x}{\eta}) \Phi(\eta,\mu).
\eqno{(14)}
$$

Substituting (14) in (11), we obtain the characteristic equation
$$
(\eta-\mu)\Phi(\eta,\mu)=\dfrac{\eta}{2}K_0n^{(0)}(\eta)+\dfrac{3}{2}\mu
\eta K_1 n^{(1)}(\eta),
$$
where
$$
n^{(k)}(\eta)=\int\limits_{-1}^{1}\mu^k \Phi(\eta,\mu)d\mu, \qquad
k=0,1.
$$

It is obvious, that
$$
n^{(1)}(\eta)=\eta(E-K_0)n^{(0)}(\eta),
$$
where $E$ is the unit  matrix of the second order.

We obtain now the characteristic equation
$$
(\eta-\mu)\Phi(\eta,\mu)=\dfrac{1}{2}\eta
D(\mu\eta)n(\eta),
\eqno{(15)}
$$
$$
n(\eta)\equiv n^{(0)}(\eta)=\int\limits_{-1}^{1}\Phi(\eta,\mu)\,d\mu,
$$
where
$$
D(\mu \eta)= K_0+3\mu\eta K_1(E-K_0)=$$$$=\left[
  \begin{array}{cc}
    3g\mu\eta & 0 \\
   \dfrac{g_1}{g_2} & 1
  \end{array}\right],
  $$
  $$
  g=1-\dfrac{g_1^2}{g_0g_2}=0.96288.
$$

For $\eta \in (-1,1)$, we use Eqs. (15) in the class of
generalized functions \cite{Vlad} to find the eigenvectors
$$
\Phi(\eta,\mu)=F(\eta,\mu)n(\eta)
$$
of the continuous spectrum, where
$$
F(\eta,\mu)=\dfrac{1}{2}\eta D(\mu  \eta)P \dfrac{1}{\eta-\mu}+
\Lambda(\eta)\delta(\eta-\mu),
$$
$$
n(\eta)=\left[
  \begin{array}{c}
    n_1(\eta) \\n_2(\eta)
  \end{array}\right] =\int\limits_{-1}^{1}\Phi(\eta,\mu)\,d\mu,
\eqno{(16)}
$$
$F(\eta,\mu)$  is the eigen matrix--function, the symbol $P x^{-1}$ denotes
the principal value of the integral in the integration
of  $x^{-1}$, $\delta(x)$ is the delta function, and
$\Lambda(z)$ is the dispersion matrix,
$$
\Lambda(z)=E+  \dfrac{z}{2}
\int\limits_{-1}^{1} \dfrac{D(\mu z)}{\mu-z}d\mu.
$$

The dispersion matrix has following elements
$$
\Lambda_{11}(z)=1+3gz^2\lambda_0(z)\equiv \lambda_1(z),
$$
$$
\Lambda_{12}(z) \equiv 0,
$$
$$
\Lambda_{21}(z)=\dfrac{g_1}{g_2}\lambda_0(z)-\dfrac{g_1}{g_2},
$$
$$
\Lambda_{22}(z)=\lambda_0(z).
$$

We represent the dispersion matrix in the explicit
form
$$
\Lambda(z)=\lambda_0(z)D(z^2)+D_0,
$$
where
$$
D_0=\left[ \begin{array}{cc}
1&0\\-\dfrac{g_1}{g_2} &  0\end{array}\right],
$$
$$
\lambda_0(z)=1+\dfrac{z}{2} \int\limits_{-1}^{1} \dfrac{du}{u-z}.
$$

Its determinant, called the dispersion function, is given by
$$
\lambda(z)=\det \Lambda(z)= \lambda_0(z)\lambda_1(z).
$$

\section{5. Eigenvectors of the discrete spectrum}

By definition \cite{Lat2}, the set of zeros of the dispersion
function is called the discrete spectrum.

The function $\lambda_0(z)$ has a single double zero at
the point at infinity $z_i=\infty$. Two eigensolutions
of Eq. (11) correspond to this zero
(they coincide with the eigenvectors of the characteristic equation):
$$
h_0(x,\mu)=\mu \left[
  \begin{array}{c}
    1 \\ 0 \end{array}\right], \qquad
h_1(x,\mu)=\left[
  \begin{array}{c}
    0 \\ 1 \end{array}\right].
$$

We find zeros of $\lambda_1(z)$ outside the cut $[-1,1]$.
It is obvious that $\lambda_1(\infty)=1-g=0.03712>0$.
We take the contour $\gamma_\varepsilon$  (see Fig. 1)
that encompasses the cut
$[-1,1]$ at a distance $\varepsilon$ from it such that there
are no zeros of the function $\lambda_1(z)$ inside the contour.
The number $N$ of zeros of $\lambda_1(z)$ outside the contour is
equal to the increment of the argument of this function
according to the argument principle \cite{Gah}, i.e.,
$$
N=(2\pi)^{-1}\Delta_{\gamma_\varepsilon}\arg \lambda_1(z),
$$
where the symbol $\Delta_{\gamma_\varepsilon}f$
means the increment of {$f$} along $\gamma_\varepsilon$.
Passing to the limit as $\varepsilon \to 0$ in this equality, we obtain
$$
N= \dfrac{1}{2\pi}\Delta_{(-1,1)}\arg \dfrac{\lambda_1^+(\mu)}
{\lambda_1^-(\mu)},
$$
where {\ $\lambda_1^{\pm}(\mu)$} are the boundary values
of $\lambda_1(z)$ on the interval
$(-1,1)$ from above and from below,
$$
\lambda_1^{\pm}(\mu)=
\lambda_1(\mu)\pm 3i\pi g\mu^3.
$$

\begin{figure}[h]
\begin{center}
\includegraphics[width=10cm, height=14cm]{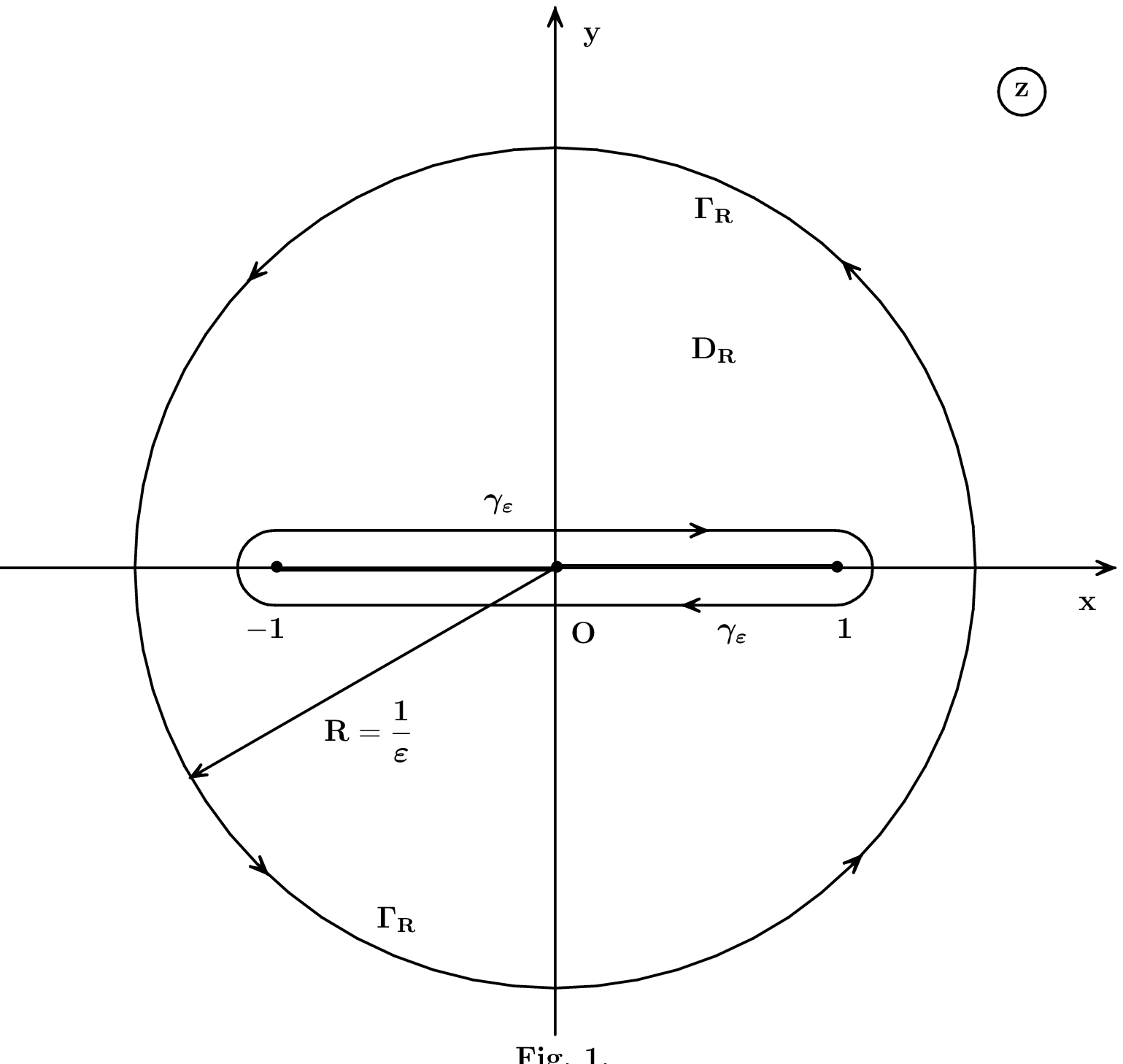}
\end{center}
\begin{center}
Fig. 1.   Dependence of the temperature jump coefficient
on the parameter 7: the specular reflection coefficient is
q = 0.3 for curve 1,
q = 0.5 for curve 2, and q = 0.8 for curve 3.
\end{center}
\end{figure}

Taking into account that $\rm Re\, \omega^+(\mu)=\omega(\mu)$
is an even function and
$\rm Im\, \omega^+(\mu)$  is odd, we have
$$
N=\dfrac{1}{\pi}\Delta_{(0,1)}\arg G(\mu),
$$
$$
G(\mu)=\dfrac{\lambda_1^+(\mu)}{\lambda_1^-(\mu)}.
$$

Let $\theta_1(\mu)=\arg \lambda_1^+(\mu)$ be the principle
value of the argument specified by the
condition $\theta_1(0)=0$. Because
$$
\overline{\lambda_1^+(\mu)}=
\lambda_1^-(\mu),\; |\lambda_1^+(\mu)|=|\lambda_1^-(\mu)|,
$$
we have $\arg G(\mu)=2\theta_1(\mu)$ and therefore
$$
N= \dfrac{2}{\pi} \Delta_{(0,1)}\theta_1(\mu).
$$

It is easy to see that the angle
$\theta_1(\mu)$ on the cut $[0,1]$ has an increment equal to $\pi$;
consequently, we have $N = 2$ (the number of zeros is
equal to two). We let $\pm \eta_0 \quad(\eta_0=1.27573)$ denote
these zeros. In view of the equality $\lambda_1(z)=\lambda_1(\bar
z)$, these zeros are real.
Two eigensolutions $h_{\pm \eta_0}(x,\mu)$, where
$$
h_{\eta_0}(x,\mu)=\Phi(\eta_0,\mu)\exp(-\dfrac{x}{\eta_0}),
\eqno{(17a)}
$$
$$
\Phi(\eta_0,\mu)=\dfrac{\eta_0}{2}\dfrac{D(\mu\eta_0)}{\eta_0-\mu}
n(\eta_0),
\eqno{(17b)}
$$
correspond to them.

We note that the homogeneous equation
$$
\Lambda(\eta_0)n(\eta_0)=\bf 0
$$
has a nonzero solution because
$$
\det \Lambda(\eta_0) \equiv \lambda
(\eta_0)=0.
$$

We represent Eq. (17) in the form of two scalar equations
$$
\lambda_1(\eta_0)n_1(\eta_0)=0,
\eqno{(18)}
$$
$$
\dfrac{g_1}{g_2}[\lambda_0(\eta_0)-1]n_1(\eta_0)+
\lambda_0(\eta_0)n_2(\eta_0)=0.
\eqno{(19)}
$$

In view of the condition $\lambda_1(\eta_0)=0$,
it follows from Eq. (18) that the upper element $n_1(\eta_0)$
is arbitrary and nonzero.
From Eq. (19), we now find the lower element of the vector
$n(\eta_0)$,
$$
n_2(\eta_0)=-\dfrac{g_1}{g_2}
\Big(1- \dfrac{1}{\lambda_0(\eta_0)}\Big)n_1(\eta_0).
$$

From the equation
$$
\lambda_1(\eta_0)=1+3g\eta_0^2\lambda_0(\eta_0)=0
$$
we find
$$
\lambda_0(\eta_0)=-\dfrac{1}{3g\eta_0^2}.
$$

Hence, the vector $n(\eta_0)$ has the form
$$
n(\eta_0)=\left[
  \begin{array}{c}
    1 \\-\dfrac{g_1}{g_2}(1+3g\eta_0^2)  \end{array}\right]
    n_1(\eta_0).
$$

We substitute this vector in the second condition in (17) and obtain
$$
\Phi(\eta_0)=\dfrac{1}{\eta_0-\mu}\left[
  \begin{array}{c}
    \mu \\ -\dfrac{g_1}{g_2}\eta_0
      \end{array}\right],
$$
for
$$
n_1(\eta_0)=\dfrac{2}{3g\eta_0^2}=-2\lambda_0(\eta_0).
$$

\section{6. Decomposition of the solution into eigenvectors:
The Riemann --- Hilbert boundary value problem}

We show that the solution of boundary value problem (11)--(13)
can be represented in the form of a decomposition, namely,
$$
h(x,\mu)=h_{as}(x,\mu)+A_0\exp(-\dfrac{x}{\eta_0})\Phi(\eta_0,\mu)+
$$
$$
+\int\limits_{0}^{\infty} \exp(-\dfrac{x}{\eta})F(\eta,\mu)A(\eta)\,d\eta,
\eqno{(20)}
$$
where the unknowns are the constants $\varepsilon_T $
and $A_0$ and a vector function $A(\eta)$ with the elements
$A_j(\eta),j=1,2,3$. Decomposition (20) can be represented in the form
$$
h(x,\mu)=h_{as}(x,\mu)+ A_0\exp(-\dfrac{x}{\eta_0})\Phi(\eta_0,\mu)+
$$
$$
+\exp(-\dfrac{x}{\mu})\Lambda(\mu)A(\mu)\theta_+(\mu)+
$$
$$
+\dfrac{1}{2}\int\limits_{0}^{1}\exp(-\dfrac{x}{\eta})\eta D(\mu\eta)A(\eta)
\dfrac{d\eta}{\eta-\mu},
\eqno{(21)}
$$
where
$$
\theta_+(\mu)=1, \mu \in (0,1);\quad \theta_+(\mu)=0,
\mu \notin (-1,0).
$$

Substituting decomposition (21) in boundary condition (12),
we obtain a singular integral equation with the Cauchy kernel \cite{Gah}
$$
h_{as}(0,\mu)+A_0\Phi(\eta_0,\mu)+\Lambda(\mu)A(\mu)+
$$
$$
+\dfrac{1}{2}\int\limits_{0}^{1}
 \eta D(\mu\eta)A(\eta)\dfrac{d\eta}{\eta-\mu}=
{\bf 0}, \;0<\mu<1.
\eqno{(22)}
$$

We introduce an auxiliary vector function
{\ $$
N(z)=\dfrac{1}{2}\int\limits_{0}^{1}
\eta D(z\eta)A(\eta)\dfrac{d\eta}{\eta-z}
\eqno{(23)}
$$}
and the matrix
$$
P(z)=\Lambda(z)D^{-1}(z^2).
$$

Using the boundary values $N(z)$, $\Lambda(z)$ and $P(z)$
and the corresponding Sokhotski formulas, we reduce Eq. (22)
to an inhomogeneous vector Riemann --- Hilbert boundary value
problem
$$
P^+(\mu)[N^+(\mu)+h_{as}(0,\mu)+A_0\Phi(\eta_0,\mu)]=
$$
$$
=P^-(\mu)[N^-(\mu)+h_{as}(0,\mu)+A_0\Phi(\eta_0,\mu)],\;0<\mu<1.
\eqno{(24)}
$$

Let's lead reduction to the diagonal form of the problem (25).
For this purpose we will lead to a diagonal kind the matrix
$$
P(z)=\left[ \begin{array}{cc}
\dfrac{\lambda_1(z)}{3gz^2}  & 0 \\
-\dfrac{g_1}{3gg_2z^2}&\lambda_0(z)
\end{array}\right].
$$

The matrix bringing the matrix $P(z)$ to a diagonal kind, is that
$$
S=\left[ \begin{array}{cc}
0 & -1 \\
1&\dfrac{g_1}{g_2}\end{array}\right],\quad \det S=1,\quad
S^{-1}=\left[ \begin{array}{cc}
\dfrac{g_1}{g_2} & 1 \\
-1&0\end{array}\right].
$$

It is obvious that
$$
S^{-1}P(z)S\equiv \Omega(z)={\rm diag}\Big\{\dfrac{\lambda_1(z)}{3gz^2},
\;\lambda_0(z)\Big\}.
$$

We first solve the homogeneous boundary value problem
$$
P^+(\mu)X^+(\mu)=P^-(\mu)X^-(\mu), \quad 0<\mu<1,
\eqno{(25)}
$$
corresponding to (24).

Clearly, that it is necessary to search for matrix $X(z)$ in the
form
$$
X(z)=S\,{\rm diag}\Big\{U_1(z),\; U_2(z)\Big\}S^{-1}.
$$

The method for solving such problems was developed in \cite{Lan}; therefore,
we give the solution of problem (25) without a derivation,
$$
X(z)=\left[ \begin{array}{cc}
U_1(z)  & 0 \\
\dfrac{g_1}{g_2}[U_0(z)-U_1(z)]&U_0(z)
\end{array}\right],
$$
where
$$
U_0(z)=z \exp(-V_0(z)),
$$
$$
U_1(z)=z \exp(-V_1(z)),
$$
$$
V_0(z)=\dfrac{1}{\pi}\int\limits_{0}^{1} \dfrac{\zeta_0(u)\,du}{u-z},
$$
$$
V_1(z)=\dfrac{1}{\pi}
\int\limits_{0}^{1} \dfrac{\zeta_1(u)\,du}{u-z},
$$
$$
\zeta_0(u)=\theta_0(u)-\pi,\quad\;\zeta_0(0)=-\pi,\quad\;
\zeta_0(1-0)=0,
$$
$$
\zeta_1(u)=\theta_1(u)-\pi, \quad\;\zeta_1(0)=-\pi,\;\quad
\zeta_1(1-0)=0.
$$

In this case, the angles have the form
$$
\zeta_0(u)=-\dfrac{\pi}{2}-\arctg\dfrac{2\lambda_0(u)}{\pi u},
$$
$$
\zeta_1(u)=-\dfrac{\pi}{2}-\arctg\dfrac{2\lambda_1(u)}{3g\pi u^3}.
$$

We now return to the solution of inhomogeneous problem (24).
Substituting the matrix $P^+(\mu)$ found using (25) in (24),
we obtain the problem of determining an analytic vector
function from its jump
$$
\Big[X^+(\mu)\Big]^{-1}\Big[N^+(\mu)+h_{as}(0,\mu)+
A_0\Phi(\eta_0,\mu)\Big]=$$
$$
=\Big[X^-(\mu)\Big]^{-1}
\Big[N^-(\mu)+h_{as}(0,\mu)+A_0\Phi(\eta_0,\mu)\Big].
\eqno{(26)}
$$

We note that
$$
 X^{-1}(z)=\left[
  \begin{array}{cc}
    \dfrac{1}{U_1(z)} & 0 \\
    \dfrac{g_1}{g_2}\Big(\dfrac{1}{U_0(z)}-\dfrac{1}{U_1(z)}\Big) &
    \dfrac{1}{U_0(z)}
  \end{array}\right],
$$
$$
  N(z)=\left[
  \begin{array}{c}
    N_1(z) \\
    N_2(z)
  \end{array}\right].
$$

The asymptotic behavior of these functions
in a neighborhood of the point at infinity can be described as
$$
X^{-1}(z)\sim \left[
  \begin{array}{cc}
    z^{-1} & 0 \\
    z^{-2} & z^{-1}\end{array}\right],\quad z\to \infty,
$$
$$
N(z)=\left[
  \begin{array}{c}
    N_1^{(0)}\\0  \end{array}\right]+o(1), \quad z\to \infty,
$$
where
$$
N_1^{(0)}=-\dfrac{3g}{2}
\int\limits_{0}^{1}\eta^2A_1(\eta)\,d\eta
$$
according to (23).

Taking the behavior of $X^{-1}(z)$ and $N(z)$ at
finite points of the complex plane and in the neighborhood of the point at
infinity into account, we obtain a general solution of problem (26)
$$
N(z)=-h_{as}(0,z)-A_0\Phi(\eta_0,z)+$$$$+X(z)\bigg[
    \Big[
  \begin{array}{c}
    C_1 \\0  \end{array}\Big]+\dfrac{1}{\eta_0-z}
  \Big[
  \begin{array}{c}
    d_1 \\d_2  \end{array}\Big] \bigg]  ,
$$
or, in scalar form,
$$
N_1(z)=-Bz-\dfrac{A_0z}{\eta_0-z}+U_1(z)\Big[C_1+
\dfrac{d_1}{\eta_0-z}\Big],
\eqno{(27)}
$$
$$
N_2(z)=-\varepsilon_T+\dfrac{g_1}{g_2}\dfrac{A_0\eta_0}{\eta_0-z}
+$$$$+\dfrac{g_1}{g_2}\Big[U_0(z)-U_1(z)\Big]\Big[C_1+\dfrac{d_1}
{\eta_0-z}\Big]+\dfrac{d_2U_0(z)}{\eta_0-z}.
\eqno{(28)}
$$

Here $C_1$,\; $d_1$ and $d_2$  are constants that can be found
from the solvability conditions for the boundary value problem.

We note that
$$
U_0(z)=z-V_0^{-1}+o(1), \qquad z\to \infty,
$$
$$
U_1(z)=z-V_1^{-1}+o(1), \qquad z\to \infty,
$$
where
$$
V_1^{(-1)}=-\dfrac{1}{\pi}\int\limits_{0}^{1}\zeta_1(u)\,du=0.84188,
$$
$$
V_0^{(-1)}=-\dfrac{1}{\pi}\int\limits_{0}^{1}\zeta_0(u)\,du=0.71045,
$$

Eliminating the pole of the function $N_1(z)$ given by
equality (27) at the point at infinity, we obtain $C_1 = B$.
Equating the limits on the right and on the left at the
point $z = \infty$ in equality
(27) for $N_1(z)$ and using (23), we obtain the equation
$$
\dfrac{3g}{2}\int\limits_{0}^{1}\eta^2A_1(\eta)\,d\eta=-A_0+
d_1+BV_1^{(-1)}.
\eqno{(29)}
$$

From the equation  $N_2(\infty)=0$, where the function $N_2(z)$
is given by equality (28), we find
$$
\varepsilon_T=-d_2+B\dfrac{g_1}{g_2}\Big[V_1^{(-1)}-V_0^{(-1)}\Big].
\eqno{(30)}
$$

Eliminating the poles of the functions  $N_1(z)$ and $N_2(z)$
at the point $\eta_0$, we obtain
$$
d_1=\dfrac{A_0\eta_0}{U_1(\eta_0)}
$$
and
$$
d_2=-\dfrac{g_1}{g_2}d_1=-\dfrac{g_1}{g_2}\dfrac{A_0\eta_0}
{U_1(\eta_0)}.
$$

Taking these relations into account, we can rewrite equality (30) as
$$
\varepsilon_T=\dfrac{g_1}{g_2}\Big[\dfrac{A_0\eta_0}{U_1(\eta_0)}
+B(V_1^{(-1)}-V_0^{(-1)})\Big].
\eqno{(31)}
$$

From definition (23) of the function $N(z)$, we have
$$
N_1(z)=\dfrac{3g}{2}z\int\limits_{0}^{1}
\dfrac{\eta^2A_1(\eta)\,d\eta}{\eta-z},
$$
whence
$$
N_1^+(\mu)-N_1^-(\mu)=3g\pi i \mu^3 A_1(\mu)
\eqno{(32)}
$$
according to the Sokhotsli formulas.

Substituting solution (27) in (32), we find
$$
\dfrac{3g}{2}\,z\,\eta^2A_1(\eta)=$$$$+\dfrac{1}{2\pi i}\Big[
\Big(C_1+\dfrac{d_1}{\eta_0}\Big)\dfrac{1}{\eta}-
\dfrac{d_1}{\eta_0(\eta-\eta_0)}\Big]
\Big[U_1^+(\eta)-U_1^-(\eta)\Big].
\eqno{(33)}
$$

Integrating equality (33) over $\eta$ from $0$ to $1$, we obtain
$$
\dfrac{3g}{2}\int\limits_{0}^{1}\eta^2A_1(\eta)d\eta=
\Big(C_1+\dfrac{d_1}{\eta_0}\Big)J(0)-\dfrac{d_1}{\eta_0}
J(\eta_0),
\eqno{(34)}
$$
where
$$
J(z)=\dfrac{1}{2\pi i}\int\limits_{0}^{1}\dfrac{U_1^+(\tau)-
U_1^-(\tau)}{\tau-z}d\tau.
$$

We give the integral representation
of the function $U_1(z)$ without a derivation,
$$
U_1(z)-z+V_1^{(-1)}=
$$
$$
=\dfrac{1}{2\pi i}\int\limits_{0}^{1}
\dfrac{U_1^+(\tau)-U_1^-(\tau)}{\tau-z}d\tau, \quad z\notin [0,1].
$$

According to this representation, we have
$$
J(\eta_0)=U_1(\eta_0)-\eta_0+V_1^{(-1)}
$$
and
$$
J(0)=U_1(0)+V_1^{(-1)}.
$$

Substituting these equalities in (34), we obtain the equation
$$
\dfrac{3g}{2}\int\limits_{0}^{1}\eta^2A_1(\eta)d\eta=
\Big(C_1+\dfrac{d_1}{\eta_0}\Big)\Big(U_1(0)+V_1^{(-1)}\Big)-$$$$-
\dfrac{d_1}{\eta_0}\Big(U_1(\eta_0)-\eta_0+V_1^{(-1)}\Big).
\eqno{(35)}
$$

From Eqs. (29) and (35), we now obtain
an equation from which we find
$$
A_0=-BU_1(\eta_0).
$$

Substituting the
found value $A_0$ in (31), we find that the temperature jump is given by
$$
\varepsilon_T=B\dfrac{g_1}{g_2}\Big[\eta_0+V_1^{(-1)}-
V_0^{(-1)}\Big].
\eqno{(36)}
$$

\section{7. Temperature jump}

We express the temperature jump in terms of the heat flux,
which is only transferred by the normal component and is
proportional to its velocity \cite{H}. We note that the mean
velocity of the gas (the normal component and Bose condensate)
is equal to zero in the gas volume. We calculate the heat flux using the
formula
$$
{\bf Q}= \int f {\bf v} \dfrac{mv^2}{2}\,\dfrac{(2s+1)d^3p}
{(2\pi\hbar)^3}, \quad {\bf p}=m{\bf v}.
\eqno{(37)}
$$

Passing to dimensionless quantities in the integral in (37), we obtain
$$
{\bf Q}= \dfrac{(2s+1)m^4}{2(2\pi\hbar)^3\beta_s^6}
\int [f_B^s+\varphi(x,{\bf C}) g(C)]{\bf C} C^2\,d^3C,
%\eqno{(6.2)}
$$
or
$$
{\bf Q}=\dfrac{(2s+1)mk^3T_s^3}{2\pi^3\hbar^3}\times
$$
$$
\times\int
\Big[Ch_1(x,\mu)+C^2h_2(x,\mu)\Big]{\bf C} C^2g(C)d^3C.
%\eqno{(6.2)}
$$

Taking into account that
$$
{\bf C}=(C\mu,C\sin \theta\cos\chi,C\sin \theta\sin \chi),
$$
$$
d^3C=C^2d\mu dCd\chi,
$$
we obtain
$$
Q_x(x)=$$$$=Q_0\int\limits_{-1}^{1}\int\limits_{0}^{\infty}
\Big[Ch_1(x,\mu)+C^2h_2(x,\mu)\Big]\mu C^5 g(C)\,d\mu dC,
$$
where
$$
Q_0=\dfrac{(2s+1)mk^3T_s^3}{\pi^2\hbar^3}.
$$

We calculate the inner integral in (38),
$$
Q_x=Q_0\Big(g_1 \int\limits_{-1}^{1}\mu h_1(x,\mu)d\mu+
g_2 \int\limits_{-1}^{1}\mu h_2(x,\mu)d\mu\Big).
$$

Taking into account that the heat flux is conserved
(i.e.,  $Q_x=\const$), we find
$$
Q_x=Q_0 g_1\dfrac{2}{3}B,
$$
hence
$$
B=\dfrac{3Q_x}{2Q_0g_1}.
\eqno{(39)}
$$
Taking (39) into account, we represent (36) in the form
$$
\Delta T=R Q_x,
\eqno{(40)}
$$
where $R$ is the Kapitsa resistance,
$$
R=\dfrac{3}{2g_2}\Big[\eta_0+V_1^{(-1)}-V_0^{(-1)}\Big]
\dfrac{\pi^2\hbar^3}{(2s+1)mk^3T_s^2},
%\eqno{(6.6)}
$$
or
$$
R=0.58514\dfrac{\pi^2\hbar^3}{(2s+1)mk^3T_s^2}=$$$$=
5.77510\dfrac{\hbar^3}{(2s+1)mk^3T_s^2}.
\eqno{(41)}
$$

Formula (40) is the sought temperature jump
(the Kapitsa jump) in the degenerate Bose gas.
The coefficient $R$ of the temperature jump, called
the Kapitsa resistance, is given by (41). It follows from relation (1)
that obtained formula (41) is applicable under the condition
$$
T \geqslant\dfrac{4\pi \hbar^2 an}{mk}.
$$

\section{8. Analysis of results}

It can be seen from (41) that the Kapitsa resistance
increases without bound as the temperature $T_s$ decreases.
At present, there are no experimental data on the Kapitsa
resistance for a degenerate Bose gas. There exist only
data on the Kapitsa resistance for liquid helium
$^4\rm He$ \cite{Enz}, \cite{H}.
According to these data, there exists a divergence of the
Kapitsa resistance as $T_s\to 0$ in the case of liquid helium.
The experimental data thus agree qualitatively with formula (41).

\end{document}